# Reply to Comment on "Localized Behavior near the Zn Impurity in YBa$_2$Cu$_4$O$_8$ as Measured By Nuclear Quadrupole Resonance

G. V. M. Williams[1,2], S. Krämer[3,*], J. L. Tallon[1,2], R. Dupree[4], and J. W. Loram[5]

[1]*Industrial Research, P.O. Box 31310, Lower Hutt, New Zealand.*
[2]*The MacDiarmid Institute, Victoria University of Wellington, P.O. Box 600, Wellington.*
[3]*2. Physikalisches Institut, Universität Stuttgart, D-70550 Stuttgart, Germany.*
[4]*Department of Physics, University of Warwick, Coventry, CV4 7AL.*
[5]*IRC in Superconductivity, Cambridge University, Cambridge CB3 0HE, United Kingdom.*

26 October 2004

**ABSTRACT**

Julien *et al.* have commented on two of our publications claiming that we have made erroneous interpretations of the nuclear magnetic resonance (NMR) and nuclear quadrupole resonance (NQR) data. Specifically, they believe that their model of an extended staggered moment about a Zn impurity is the only interpretation of the data [Julien *et al.*, Phys. Rev Lett. **84**, 3422 (2000)]. Not only does their claim ignore models presented by other authors, we show that the model of Julien *et al.* [Phys. Rev Lett. **84**, 3422 (2000)] does not consistently reproduce all of the NMR data.

**Introduction**

Julien *et al.* [1] have criticized two of our publications reporting the results from NMR and NQR measurements on YBa$_2$(Cu$_{1-x}$Zn$_x$)$_4$O$_8$ [2,3]. These papers were published more that two and a half years ago and it should be accepted that the interpretation of data may change when new data or models are presented. However, we do not agree that the Julien *et al.* [4] model is the only interpretation of the data and that our model is in contradiction with the ".. facts established by the rest of the NMR community.".

We have previously interpreted our Cu NQR data in terms of a localized charge pile-up about the Zn impurity [2,3] and we used the Zn-induced local moment model of Mahajan *et al.* [5] to interpret our $^{89}$Y NMR data [6] where it was assumed that Zn induces a moment on the nearest-neighbor Cu sites. This assumption of a very localized Zn-induced local moment has also been used by other authors [7,8,9,10,11] and the appearance of localized charge is supported by tunneling measurements on Bi$_2$Sr$_2$Ca(Cu$_{1-x}$Zn$_x$)$_2$O$_8$ [12]. Julien *et al.* assume antiferromagnetically correlated spins on the Cu sites out to at least 5 lattice parameters away from the Zn impurity as well as "enhanced antiferromagnetic spin correlations" [4].

Our localized charge pile-up model was based on a Zn-induced Cu NQR satellite peak, which we suggested may be associated with the localized spin assumed from $^{89}$Y NMR measurements. Julien *et al.* [1] claim that attributing the lower NQR frequency from the Zn-induced satellite peak to a lower hole concentration is incorrect because of possible "lattice" effects. This ignores the experimental data that show a strong correlation between the hole concentration and the Cu NQR frequency [13-16]. We do not believe that the results from Cu NQR measurements on a antiferromagnetic La$_2$CuO$_4$ sample doped with Zn [18] proves that our use of the experimental correlation between the Cu NQR frequency and hole concentration is invalid as suggested by Julien *et al.* [1]. This correlation is supported by a recent theoretical study that showed that the Cu electric field gradient is dominated by large positive and negative effects arising from the Cu 3d orbitals and virtual hoping from the neighboring oxygen ions to the unoccupied Cu 4d orbitals [17].

Our Cu spin-spin relaxation data do not support the existence of the "enhanced antiferromagnetic spin correlations" assumed by Julien *et al.* [4]. We found that the value of the Gaussian component, $^{63}T_{2G}$, from Cu sites that are not nearest-neighbor to the Zn impurity is comparable to that observed in pure YBa$_2$Cu$_4$O$_8$ where it has been argued that $1/^{63}T_{2G}$ is weighted by $\chi'(\mathbf{q})$ about $\mathbf{q} = \mathbf{Q}_{AF}$, where $\mathbf{Q}_{AF}$ is the antiferromagnetic wavevector. Therefore, our data are consistent with the absence of any *enhancement* of the antiferromagnetic spin fluctuations for all sites that are not next-nearest-neighbor to the Zn impurity since $\chi'(\mathbf{Q}_{AF})$ remains *unchanged*.

We previously showed that our model is supported by the functional dependence of the spin-echo intensity on the time between the $\tau_{\pi/2}$ and $\tau_\pi$ pulses, which is different for the satellite and main peak [2]. Unfortunately Julien *et al.* have misrepresented our arguments concerning the spin-spin relaxation rate, $1/^{63}T_2$ [1]. We agree with Julien *et al.* that the interpretation of $T_{2G}$ takes only the $I_z^i I_z^j$ term into account. Consequently the equation for $T_{2G}$, which we quoted in our paper (equation 5 in ref. 2), takes only this term into account. But it should also be clear that the $I_z^i I_z^j$ term only contributes to $T_{2G}$ if *both* spins (i *and* j) are flipped by the refocusing radio-frequency-pulse (RF-pulse). Therefore they have to be within the excitation window of the RF-pulse. In our paper we addressed these spins with the terminology *like spins*. This terminology is also used by other researchers [19-21].

By considering $^{63}$Cu NMR and NQR data, $^{89}$Y NMR data as well as recent specific heat data we show below that the Julien *et al.* model [4] is not the only possible interpretation of the overall effect of Zn in $YBa_2Cu_4O_8$.

**Experimental Details**

Samples of $YBa_2(Cu_{1-x}Zn_x)_4O_8$ were synthesized by standard repeated solid-state reaction and pellets of approximately 1 g were studied by high-precision differential specific heat measurements as described elsewhere [22,23]. Although the samples were apparently free of impurity phases as indicated by x-ray powder diffraction the present specific heat measurements are extremely sensitive to magnetic impurities at low temperatures and a weak anomaly associated with residual $Y_2BaCuO_5$ was observed at 17 K on top of the bulk electronic specific heat for $YBa_2Cu_4O_8$. In addition a weak Schottky term was found at low temperature associated with free spins from residual $BaCuO_2$. The data reported below has the Schottky term, with its well-defined temperature and field dependence, removed.

**Results, Analysis and Discussion**

Julien *et al.* [1] state that comments by Itoh *et al.* [24] are "complementary" to those that they have presented. We question this statement. It should be made clear that Itoh *et al.* did not make any criticism of our work. Itoh *et al.* [24] presented a different interpretation of the Cu NQR data where they assumed a wipe-out of the Cu NQR signal near the Zn impurity and a non-unique model of the Cu spin-lattice relaxation rate. However, similar to us and Ishida *et al.* [25], they observed a distribution of Cu spin-lattice relaxation rates that was not reported by Julien *et al.* [4]. Since Julien *et al.* [1] refer later in their Comment to possible wipe-out of the Cu NQR signal near the Zn impurity, it is apparent that it is this new assumption that Julien *et al.* [1] believe is a comment on our work. This assumption was also used by Ouazi *et al.* [26] who are referenced by Julien *et al.* [1] in their Comment, in an unpublished NMR report on Li-doped $YBa_2Cu_3O_{7-\delta}$. It should be noted that a wipe-out to the spatial extent now being assumed was not mentioned in the earlier report by Julien *et al.* [4].

Itoh *et al.* [24] use a complicated statistical model and deduce that the Cu NQR signal is completely wiped-out for Cu sites that are 1$^{st}$, 2$^{nd}$ and 3$^{rd}$ nearest-neighbors from the Zn impurity. However, they later state that ".. one should be careful with the results from intensity analysis.". In the same paragraph they state that the safest conclusion is that the satellite peak comes from the Zn neighbor sites and the main peak arises from Cu sites that are not near the Zn impurity. This is the same conclusion that we had reached [2]. It should be noted that the "all-or-nothing" wipe-out model used by Itoh *et al.* [24] is unrealistic. It assumes a complete wipe-out of the Cu signal from the 1$^{st}$, 2$^{nd}$ and 3$^{rd}$ nearest neighbors to the Zn impurity and a complete recovery for the 4$^{th}$ and greater nearest neighbors. It is then assumed that some "lattice effect" causes a shift in the resonance frequency for the 4$^{th}$ nearest neighbor Cu site leading to a satellite peak, which is not experienced by the 5$^{th}$ and greater nearest-neighbors. Itoh *et al.* [24] are aware of the problems with this model. Specifically, a wipe-out of the Cu NQR intensity should be accompanied by a more gradual recovery of the Cu NQR intensity for more-distant Cu sites. Thus, the spin-spin decay should be more rapid for the satellite peak. Unfortunately, this is not observed [2]. Rather, the spin-spin decay is slower for the satellite peak, which we have shown is consistent with the satellite peak arising from the Cu sites that are nearest-neighbor to the Zn impurity.

We have performed new Cu NQR measurements on YBa$_2$(Cu$_{1-x}$Zn$_x$)$_4$O$_8$ for Zn concentrations of up to 10 % per planar Cu site [27]. In the model of Itoh *et al.* [24] the total intensity for the 10 % Zn per planer Cu sample should be reduced to ~20 % of that in the pure compound, which is not observed. As we have mentioned previously [2], the ratio of the integrated intensity from the satellite to the main peak is consistent with the satellite peak arising from the Cu sites that are nearest-neighbor to the Zn impurity. This is evident in figure 1 where this ratio is plotted for our samples (solid circles) and the reported results from Itoh *et al.* (open circles [24]). The expected ratios were calculated by randomly placing Zn on a 400×400 lattice to represent the CuO$_2$ plane. We find that our data and the data of Itoh *et al.* are consistent with the satellite peak arising from Cu sites that are nearest-neighbor to the Zn impurity (solid curve). However, the model suggested by Itoh *et al.* (dashed curve) does not fit the experimental data. It is clear that our model provides a better description of the data.

Julien *et al.* [4] have suggested that Zn induces "*enhanced* antiferromagnetic correlations" leading to enhanced antiferromagnetic spin fluctuations and an increase in $1/^{63}T_1T$ at low temperatures, where $1/^{63}T_1$ is the Cu spin-lattice relaxation rate. We show below that our $1/^{63}T_1T$ data, as well as new electronic heat capacity data are consistent with a partial filling in of the normal-state pseudogap and no enhanced antiferromagnetic spin fluctuations.

Our Cu spin-lattice relaxation rate data from NQR measurements on YBa$_2$(Cu$_{1-x}$Zn$_x$)$_4$O$_8$ are plotted above T$_c$ in figure 2a as $^{63}T_1T$, along with data for pure YBa$_2$Cu$_4$O$_8$ from other researchers [28-30]. We have previously shown that $^{63}T_1T$ from YBa$_2$Cu$_4$O$_8$ can be modeled by $^{63}T_1T = a_0(T+\vartheta)/\chi_s(T)$ where $\chi_s(T)$ is the static spin susceptibility, and $\vartheta = 0$ [31]. While this relation can be derived from the Millis, Monien and Pines (MMP) model [32], it represents the assumption used in the NMR community that the antiferromagnetic spin fluctuation spectrum increases with decreasing temperature in the superconducting cuprates. The increase in $^{63}T_1T$ at low temperatures arises from the normal-state pseudogap. It has been argued that the normal-state pseudogap exists in both the spin and charge spectrum [33]. In the absence of the normal-state pseudogap $^{63}T_1T$ is expected to follow the dashed line in figure 2a. Within this interpretation, if there is an enhancement of the antiferromagnetic spin fluctuations then the experimental data from YBa$_2$(Cu$_{1-x}$Zn$_x$)$_4$O$_8$ should lie below that of the dashed line. However, this does not occur and the data can be accounted for by a partial filling in of the normal-state pseudogap.

Since $\vartheta = 0$ for YBa$_2$Cu$_4$O$_8$, it is possible from the relation $^{63}T_1T = a_0(T + \vartheta)/\chi_s(T)$ to deduce the temperature dependence of $\chi_s(T)$, which is proportional to $1/^{63}T_1$. For this reason, $1/^{63}T_1$ is shown in figure 2b for temperatures above T$_c$ for YBa$_2$(Cu$_{1-x}$Zn$_x$)$_4$O$_8$. It is clear that the effect of Zn on sites that are not nearest-neighbor to the Cu sites averages to an effective partial filling in of the normal-state pseudogap.

Zn-induced filling of the normal-state pseudogap is also evident in the data of figure 2c. Here we plot $S_{el}/T$ where $S_{el}$ is the electronic entropy for a pure YBa$_2$Cu$_4$O$_8$ sample and a sample with 4% Zn per planar Cu. The normal-state region is shown by the solid curves. $S_{el}$ is obtained from the measured electronic heat capacity, $C_{el}$, using the thermodynamic relation, $C_{el} = T(\frac{\partial S_{el}}{\partial T})$. For high temperature superconducting cuprates, it has been shown that $S_{el}/T = a_W \chi_s$ [33] where $a_W$ is Wilson's ratio for nearly free independent electrons. Thus $S_{el}/T$ is proportional to $\chi_s$. Consequently, the data in figure 2c is analogous to that in figure 2b and also shows a Zn-induced filling in of the normal-state pseudogap.

Julien *et al.* [1] claim that their model is supported by $^{89}$Y NMR measurements on YBa$_2$(Cu$_{1-x}$Zn$_x$)$_4$O$_8$ and present data from unpublished work at 50 K that we do not have access to. We show below that their model [1,4] does not describe our $^{89}$Y NMR data on YBa$_2$(Cu$_{1-x}$Zn$_x$)$_4$O$_8$ and the new more localized moment model [1] will lead to multi-peak $^{63}$Cu NMR spectra.

For this purpose we model the $^{89}$Y and $^{63}$Cu NMR spectra from various experimental publications using a Hamiltonian that comprises only magnetic terms. We start by noting that the $^{89}$Y and $^{63}$Cu NMR spectra can be constructed from the Shasty-Mila-Rice Hamiltonian [34]. This leads to a hyperfine field at each $^{89}$Y site that can be written as $h(i,j) = \sum D <s> (i+\delta i, j+\delta j, k)$ where the sum is over $\delta i, \delta j = 0,1$ and $k = 1,2$ to represent the two CuO$_2$ planes, $<s>$ is the thermal average spin, and $D$ is the $^{89}$Y hyperfine coupling constant (-0.39 T [50]). At each $^{63}$Cu site the hyperfine field can be written as $h(i,j) = A<s>(i,j) + B\sum <s>(i+\delta i, j+\delta j)$ where $\delta i, \delta j = -1,1$, $A = -33$ T, and $B = 8$ T [35] for the c-axis parallel to the applied magnetic field, c||B. Similar to Walstedt *et al.* [5],

the NMR spectra are constructed using two dimensional lattices. We use two $200 \times 200$ matrices to represent the 2 CuO$_2$ planes. The Zn atoms were randomly distributed throughout the 2 lattices and $<s>$ was calculated about each Zn atom using the models discussed below. Similar to Julien *et al.* [4] we assume that $<s>(i,j)$ is additive at each Cu site. The spectra were constructed from Gaussian's.

To check that this model can simulate the NMR data we used the same model as Walstedt *et al.* [7] and simulated their $^{63}$Cu NMR data from measurements on YBa$_2$(Cu$_{0.97}$Zn$_{0.03}$)$_3$O$_7$. They assumed a local moment on the Zn site that induces a spin density oscillation that can be written as $<s>(i,j) = \gamma(-1)^{i+j}\sum \exp(-(i^2+j^2)/(4\xi^2))$ where $\xi$ is the antiferromagnetic correlation length divided by the average ab-plane lattice parameter. The parameter $\gamma$ contains the local moment on the Zn site and can be written as $\gamma = -J(\chi_0/\mu_B^2)(\beta^{1/2}/\xi)\xi<S_L>/(16\pi)$ where $J$ is the exchange energy, $\chi_0$ is the host-band uniform susceptibility, $\mu_B$ is the Bohr magneton, $\beta$ is a scale parameter and $<S_L>$ is related to the thermal average spin. The local moment, $P_{eff}$, used by Walstedt *et al.* [7] was assumed to be 1 $\mu_B$ and $<S_L>$ can be written as $<S_L> = (P_{eff}^2 B)/(g\mu_B 3kT)$. We were able to model their $^{63}$Cu NMR linewidth data and produce similar $^{63}$Cu NMR spectra for c||B with $<s>(i,j) = -\gamma'(B/T)(-1)^{i+j}\exp(-(i^2+j^2)/(4\xi^2))$, where $\gamma' = 0.0052$ and $B = 7.5$ T. We used the same value of $\xi$ used by Walstedt *et al.* [7] ($\xi = 1.25$) and intrinsic linewidths of 0.01 T. It should be noted that for the ab-plane parallel to applied magnetic field, ab||B, $A = 4.8$ T [35]. Thus this model produces a $^{63}$Cu NMR linewidth for c||B that is larger than that for ab||B. Consequently, the statement by Julien *et al.* [4] that a larger $^{63}$Cu NMR linewidth for c||B when compared with ab||B ".. confirms without any detailed model that the staggered component of the magnetization is dominant" is not correct.

Julien *et al.* [1] assume a spatial dependence for $<s>$ which can be approximated from figure 1a of ref. [1] by,

$$<s>(i,j) = -\gamma'(B/T)(-1)^{i+j}\exp(-(i^2+j^2)^{1/2}/r_0), \qquad [1]$$

where $r_0$ is a parameter that defines the extent of their local moment. This model can qualitatively reproduce their $^{63}$Cu NMR spectra at 80 K in figure 1d of ref [4] from NMR measurements on YBa$_2$(Cu$_{0.99}$Zn$_{0.01}$)$_3$O$_{6.7}$, with $r_0$ greater than ~2.7. Julien *et al.* [4] state that the local moment per Zn is proportional to the sum of $<s>(i,j)$. They now use a more localized $<s>(i,j)$ to fit their unpublished $^{89}$Y and $^{63}$Cu NMR from NMR measurements on YBa$_2$(Cu$_{0.995}$Zn$_{0.005}$)$_4$O$_8$, where $r_0$ is ~1.53 as estimated from figure 1a of ref. [1]. While this model can lead to a broadening of the $^{63}$Cu

NMR spectra, a problem arises when it is applied to the $^{89}$Y NMR data. We [6] and Mahajan *et al.* [5] showed that 3 peaks are clearly visible in the $^{89}$Y NMR data from YBa$_2$(Cu$_{1-x}$Zn$_x$)$_3$O$_{7-\delta}$ and YBa$_2$(Cu$_{1-x}$Zn$_x$)$_4$O$_8$. This is apparent in figure 3b where the $^{89}$Y magic angle spinning (MAS) NMR spectrum at 132 K and 11.71 T from YBa$_2$(Cu$_{0.9825}$Zn$_{0.0175}$)$_4$O$_8$ is plotted [6]. However, the model of Julien *et al.* [1,4] only produces one additional peak as can be seen in figure 3b. The simulation was obtained with $\gamma' = 0.073$ and an intrinsic $^{89}$Y MAS NMR linewidth of 8 ppm. The resultant spatial dependence of $<s>(i,0)$ can be seen in figure 3a (dashed curve). Thus, while the model of Julien *et al.* [1,4] will produce broadening of the main $^{89}$Y NMR peak as reported by Mahajan *et al.*, it does not provide a good representation of our $^{89}$Y MAS NMR data. If the claim by Julien *et al.* [1] that their colleagues [5] used a similar spatial dependence of $<s>$ to fit $^{89}$Y NMR data from YBa$_2$(Cu$_{1-x}$Zn$_x$)$_3$O$_{7-\delta}$ is correct, it would mean that the analysis by Mahajan *et al.* [5] is flawed because the model does not provide a good representation of the experimental data.

We previously showed that it is possible to describe our $^{89}$Y MAS NMR data [6] using the model of Mahajan *et al.* [5], where local moments on the nearest-neighbor Cu sites are assumed. We now consider the effect of an induced local moment of the form now assumed by Julien *et al.* [1,4]. Assuming that a Zn-induced moment exists, and $<s>$ is given by equation 1, we find that our $^{89}$Y MAS NMR spectra can be modeled with $\gamma' = 0.20$, $r_0 = 0.54$ and an intrinsic linewidth of 25 ppm (solid curve figure 3b). The resultant $<s>$ is plotted in figure 3a (solid curve) at 132 K. The value of $<s>(i,0)$ for $i = 2$ is 15% of that for $i = 1$ and hence this leads to a large local moment on the four nearest-neighbor Cu sites to the Zn impurity, which is consistent with our earlier analysis [6].

The reason why the Julien *et al.* model [4] fails to reproduce our $^{89}$Y MAS NMR is due to the assumed extended nature of a local moment. This is clearly illustrated in figure 4 where the $^{89}$Y hyperfine field ratio between the nearest neighbor $^{89}$Y site and the next nearest neighbor $^{89}$Y site is plotted against $r_0$ where $<s>$ from equation 1 was used. For simplicity we consider an isolated Zn impurity and a bulk Knight shift of zero. The nearest-neighbor $^{89}$Y site leads to a satellite with the largest $^{89}$Y NMR shift and the next-nearest-neighbor $^{89}$Y site leads to a satellite closer to the main peak that is seen in the raw data in figure 3b. The experimental ratio is 2.5, while the Julien *et al.* [1] model produces a ratio near 7. A similar analysis of the data reported by Bobroff *et al.* [36] from $^{89}$Y NMR measurements on YBa$_2$(Cu$_{0.99}$Zn$_{0.01}$)$_3$O$_{6.64}$ reveals an experimental ratio of ~3.25 at 132 K. From figure 4 it can be seen that this leads to a $r_0$ of 0.56. The ratio is even lower at a temperature of ~75 K (~2.25) and implies an even more localized moment.

We have also modeled the unpublished $^{89}$Y and $^{63}$Cu NMR data from NMR measurements on YBa$_2$(Cu$_{1-x}$Zn$_x$)$_4$O$_8$ with a low Zn concentration of 1 % Zn per planar Cu site that Julien *et al.* present in figure 1 of their Comment [1]. Their $^{89}$Y NMR simulation at 14 T and 50 K can be reproduced with equation 1, their $r_0$ of 1.53, $\gamma' = 0.051$ and an intrinsic linewidth of 82 ppm as can be seen in figure 5b (dashed curve). However, it should be noted that there is no additional second peak at ~0.0065 T as claimed by Julien *et al* [1]. As we have shown above, the model of Julien *et al* [4] only produces one satellite peak and not the two peaks seen at higher Zn concentrations. The absolute value of $<s>(i,0)$ is plotted in figure 5a (dashed curve) at 14 T and 50 K.

The resultant $^{63}$Cu NMR simulation is plotted in figure 5c (dashed curve) using an intrinsic linewidth of 0.04 T but the spectrum is multi-peaked and extends out to greater than $\Delta B = \pm 0.4$ T due to the effect of spin on Cu sites that are near the Zn impurity. If it is assumed that the $^{63}$Cu NMR signal is wiped out for $(i^2 + j^2)^{1/2} \leq 3$ then it is possible to obtain simulated $^{63}$Cu NMR spectra with only one peak. However, this assumption was not made in the original paper by Julien *et al.* [1] and Itoh *et al.* only assumed wipe-out for $(i^2 + j^2)^{1/2} \leq 2$. If this is the new assumption of Julien *et al.* then their original report [4] is disputable because their assumption that their Cu NMR data proves the existence of an extended staggered moment involves an extrapolation to Cu sites for $(i^2 + j^2)^{1/2} \leq 3$ that they do not see. Furthermore, we do not observe a wipe-out of the Cu NQR signal to the extent suggested by Itoh *et al.* [24]. Therefore, the staggered moment model of Julien *et al.* [4] is not a consistent or viable interpretation of the NMR data.

It is possible to model the new unpublished data of Julien *et al.* [1] from $^{89}$Y and $^{63}$Cu NMR measurements on YBa$_2$(Cu$_{1-x}$Zn$_x$)$_4$O$_8$ with 1 % Zn per planar Cu site using a localized moment model with an additional small induced spin density oscillation. However, as we show below this also leads to multiple $^{63}$Cu NMR peaks. We use $<s> = -[\gamma_1'(B/T)(-1)^{i+j}\exp(-(i^2+j^2)^{1/2}/r_{01}) + \gamma_2'(B/T)(-1)^{i+j}\exp(-(i^2+j^2)^{1/2}/r_{02})$ where $\gamma_1' = 0.20$, $r_{01} = 0.4$, $\gamma_2' = 0.0052$, $r_{02} = 3$, an intrinsic $^{89}$Y linewidth of 82 ppm and an intrinsic $^{63}$Cu linewidth of 0.04 T. The first term represents a local moment and the second term represents a small induced spin polarization. It can be seen in figure 5b and 5c that this $<s>$ reproduces the $^{89}$Y NMR spectra (solid curve) and leads to broadened $^{63}$Cu NMR spectra (solid curve). Unfortunately, the predicted Cu NMR spectrum is again multi-peaked, which is not seen in the experimental data of Julien *et al.* [1]. Note that the spatial dependence of $<s>$ (solid curve figure 5a) is still dominated by Cu sites that are nearest-neighbor to the Zn impurity. In this model the first $^{89}$Y NMR satellite peak is to the right of the main peak. However, this peak is merged with the main peak because Julien *et al.* have presented data only at a low temperature (50 K) that is below T$_c$. While Julien *et al.* do not quote T$_c$ for their sample, Itoh *et al.* report a T$_c$ of 68 K for a sample with the same impurity content. Thus, the experimental spectrum may be broadened by flux penetration. In the model of Julien *et al.* the Zn-induced broadening follows a Curie-Weiss-like temperature-dependence and hence the spectra would be narrower at higher temperatures where the first satellite peak would then be evident in the $^{89}$Y NMR data. Finally, the spectrum would be even narrower if Julien *et al.* had performed MAS rather than static $^{89}$Y NMR measurements.

If we ignore the serious problem of the predicted multi-peaked Cu NMR spectra, it is possible to show that this local moment and small induced spin polarization model can provide a representation of our $^{89}$Y MAS NMR spectra at 132 K and plotted in figure 3b (dotted curves). In this case the magnitude of $<s>$ needs to be increased by a factor of 1.51 to enable a representation of our $^{89}$Y NMR data. It should be noted that the largest contribution to a sum of $<s>(i,j)$ about an isolated Zn impurity (~96 %) arises for $r < \sqrt{2}$ and hence within this interpretation the local moment is predominately on the nearest neighbor Cu sites. Unfortunately, this model and the model of Julien *et al.* [1,4] requires an intrinsic $^{89}$Y MAS NMR linewidth that is much larger than that found in the pure compound or at higher temperatures [6]. Thus, this model also does not provide a full description of the temperature-dependence of the $^{89}$Y MAS NMR linewidth.

Julien *et al.* [1] refer to recent unpublished work by Ouazi *et al.* [26] as providing support for their model of an extended Zn-induced staggered moment. The work to which they refer concerns Li-doped YBa$_2$Cu$_3$O$_{7-\delta}$ and involves a different interpretation of the data than that provided previously [36]. They assume a $<s>(i,j)$ involving exponentials (figure 4 inset in reference [26]) obtained from fitting their $^{17}$O NMR linewidth data. For up to 8 lattice parameters this distribution can be fitted to $<s>(i,j) \propto -[4.8(-1)^{i+j}\exp(-(i^2+j^2)^{1/2}/0.365) + (-1)^{i+j}\exp(-(i^2+j^2)^{1/2}/3.86)]$. However, this extended distribution can not fit our $^{89}$Y MAS NMR data. Furthermore, this model is different from that assumed by Julien *et al.* [1,4].

**Conclusion**

In conclusion, we have shown that comments by Julien *et al.* [1] regarding two of our papers published more that two and a half years ago are not well founded. In particular, the assertion that their extended staggered Zn-induced local moment model with a local enhancement of the antiferromagnetic spin correlations is the only possible interpretation of the data is not correct. Furthermore, their claim that their model has been "accepted for years" seriously overstates the situation. We have also shown that the increase in $1/^{63}T_1T$ at low temperatures with increasing Zn concentration in YBa$_2$Cu$_4$O$_8$ can be accounted for by a partial filling in of the normal-state pseudogap, which is supported by electronic heat capacity measurements. Finally, we have shown that their model as proposed in ref. 1 and ref. 4 is unable to reproduce the wider body of data including our $^{89}$Y MAS NMR spectra.

**Acknowledgements**

We acknowledge funding support from the New Zealand Foundation for Research Science and Technology, the New Zealand Marsden fund and the Alexander von Humboldt Foundation.

**FIGURES**

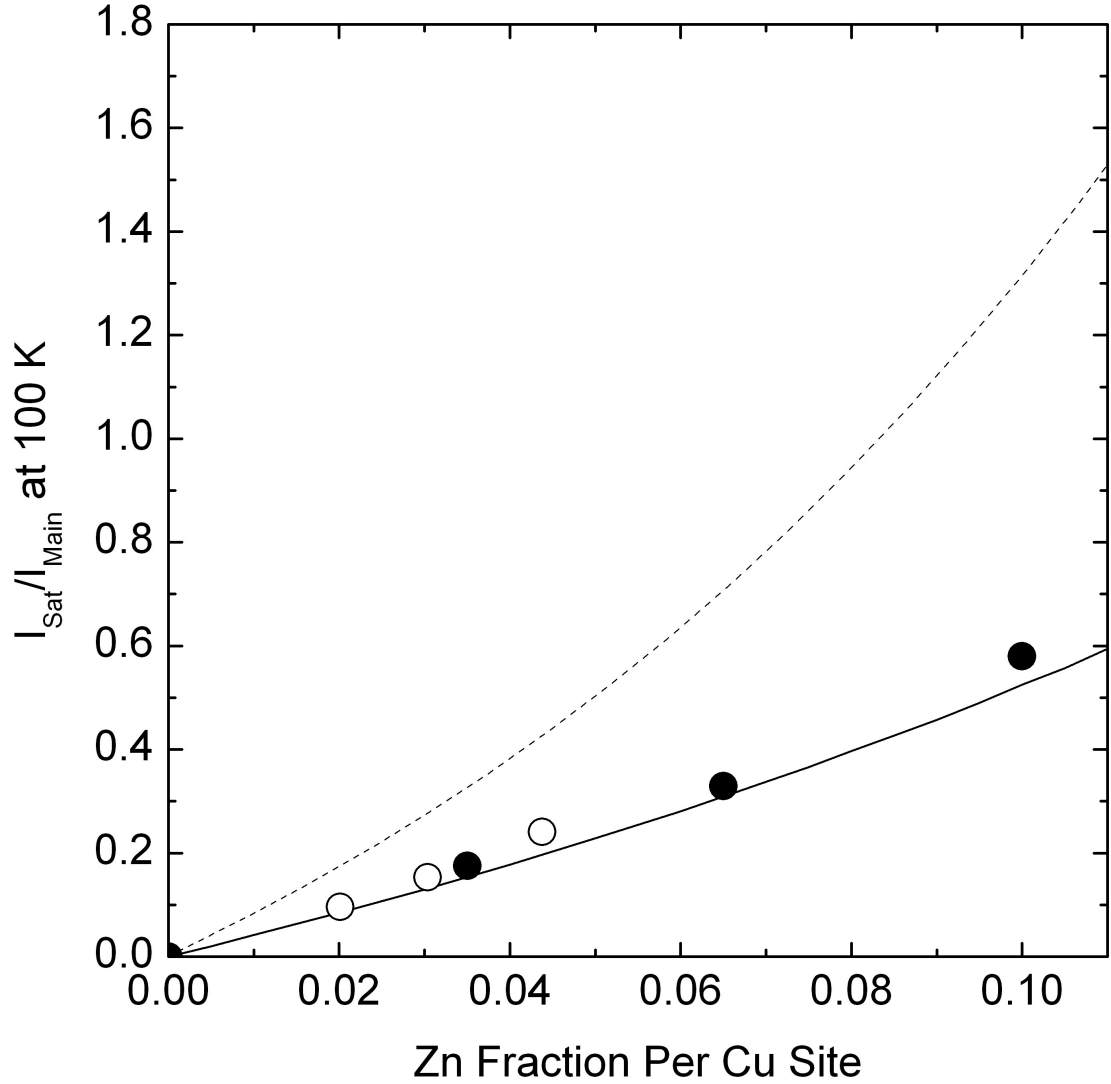

**Figure 1:** Plot of the integrated $^{63}$Cu NQR intensity of the satellite peak divided by the main peak from YBa$_2$(Cu$_{1-x}$Zn$_x$)$_4$O$_8$ against the Zn concentration per planar Cu site (2x) at 100 K (filled circles [27], open circles [24]. Also shown is the expected ratio if the satellite peak arises from Cu sites that are nearest neighbor to the Zn impurity (solid curve) or from Cu sites that are 4$^{th}$ nearest neighbor to the Zn impurity as suggested by Itoh *et al.* [24] (dashed curve).

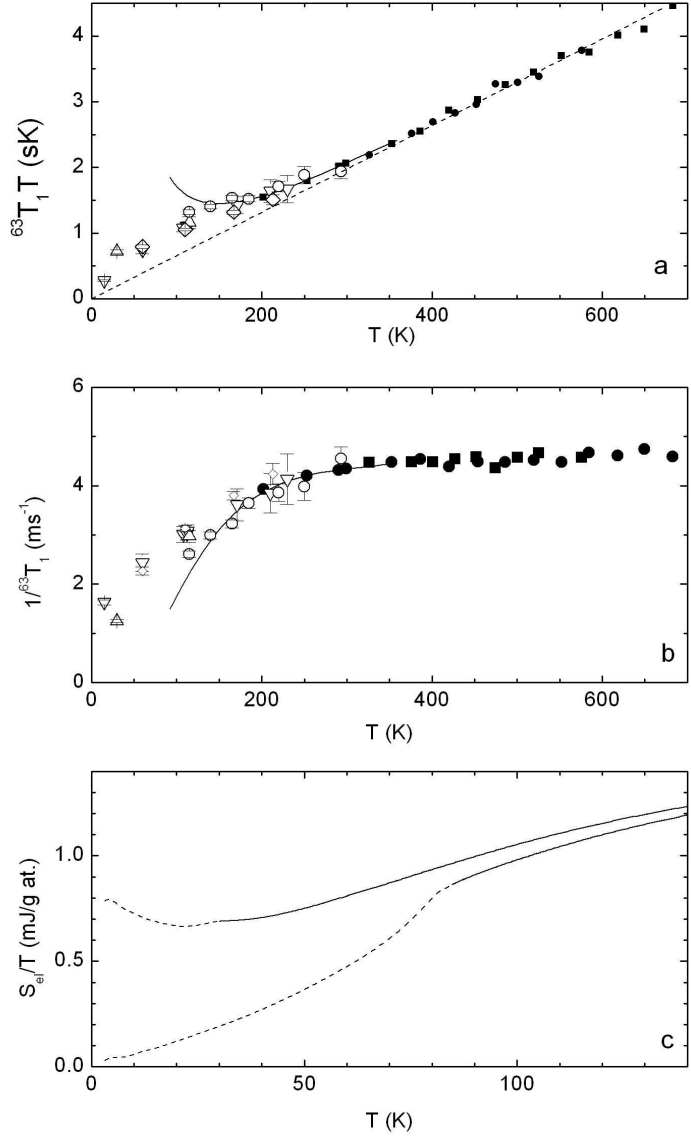

**Figure 2:** (a) Plot of $^{63}T_1T$ against temperature from the main $^{63}$Cu NQR peak for $YBa_2(Cu_{1-x}Zn_x)_4O_8$ with x=0.0875 (open circles), x=0.175 (open diamonds), x=0.25 (open up triangles) and x=3.25 (open down triangles) [2]. Also shown is data for pure $YBa_2Cu_4O_8$ (filled circles [28], filled squares [29] and solid curve [30]). The dashed line is the high temperature linear fit to the data. (b) Plot of $1/^{63}T_1$ using the same data as in (a). (c) Plot of $S_{el}/T$ against temperature for $YBa_2(Cu_{1-x}Zn_x)_4O_8$ with x=0 (lower curve) and x=0.02 (upper curve). The solid curves indicate data above $T_c$ and the dashed curves indicate date below $T_c$.

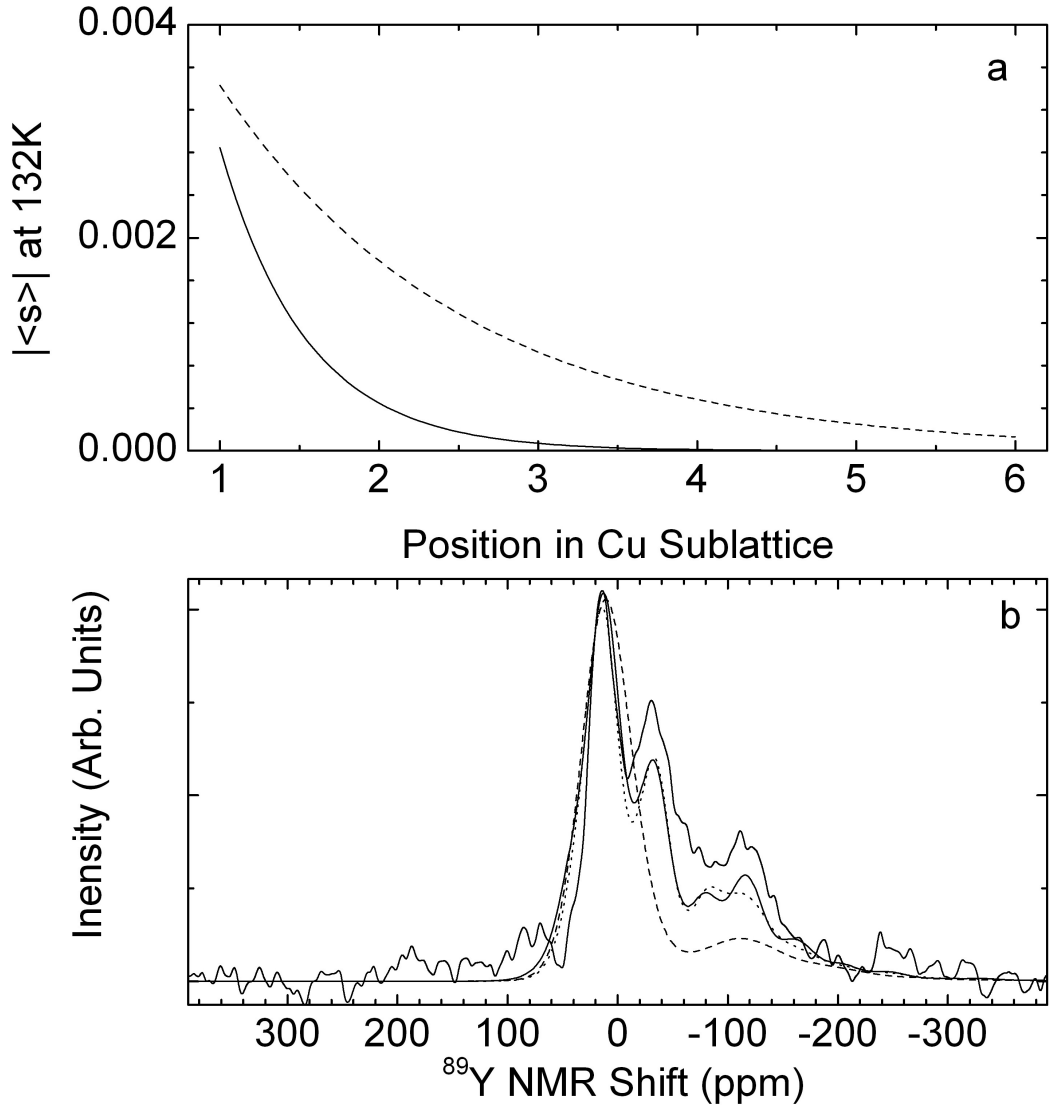

**Figure 3:** (a) Plot of the absolute value of $<s>(i,0)$ at 132 K and 11.71 T against the position in the Cu sublattice where the Zn atom is at the origin for the Julien *et al.* model [1] (dashed curve), and the local moment model (solid curve) as described in the text. The data is a one dimensional representation of the absolute value of $<s>$. (b) Plot of the $^{89}$Y MAS NMR data from YBa$_2$(Cu$_{1-x}$Zn$_x$)$_4$O$_8$ with x=0.0175 at 132 K and an applied magnetic field of 11.71 T [6]. Also shown are the spectra expected for the model of Julien *et al.* (dashed curve [1]), the local moment model (solid curve) and the local moment model with a small induced spin density oscillation (dotted curve).

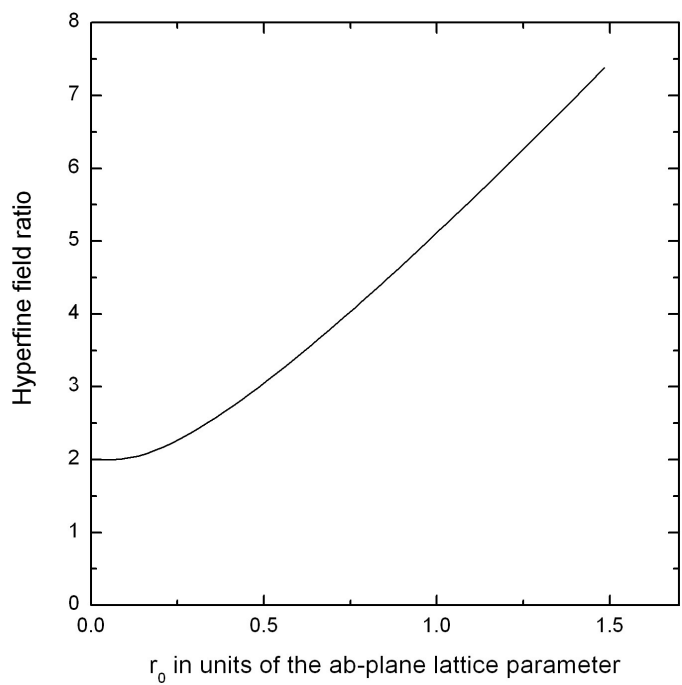

**Figure 4:** Plot of the $^{89}$Y hyperfine field from the nearest-neighbor $^{89}$Y site divided by that from the next-nearest-neighbor $^{89}$Y site against the parameter determining the extent of a Zn-induced moment as described in the text.

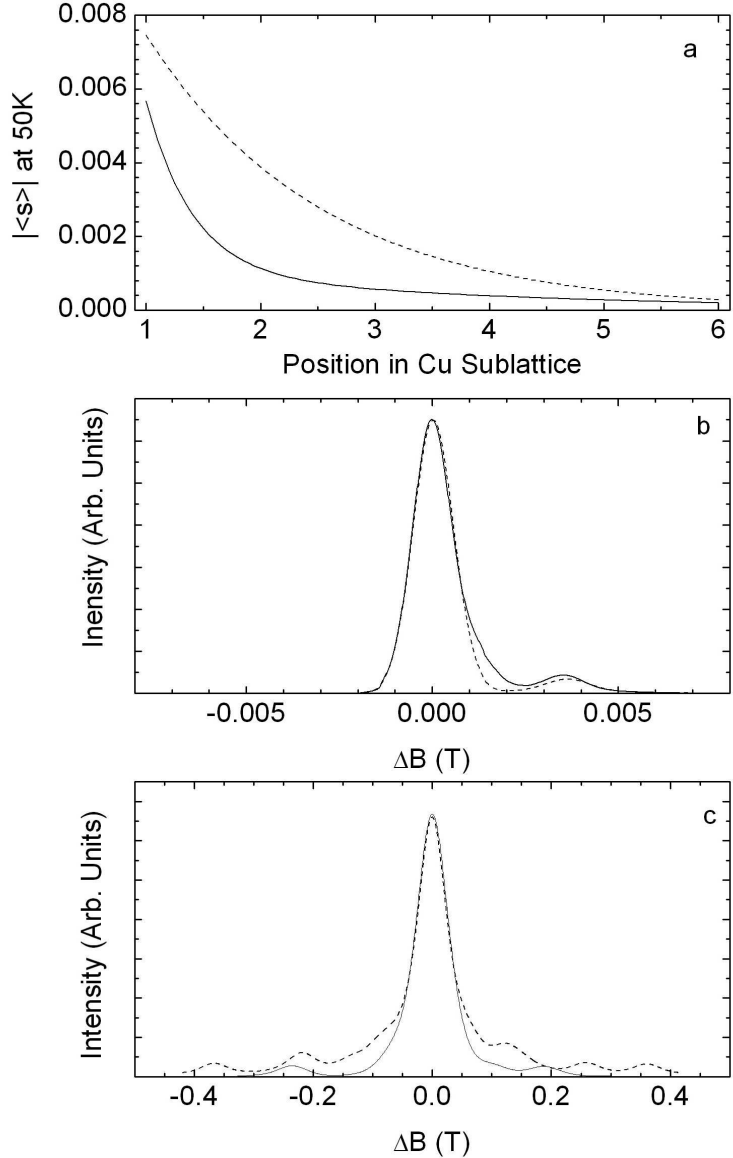

**Figure 5:** (a) Plot of the absolute value of $<s>$ at 50 K and 14 T against the position in the Cu sublattice where the Zn atom is at the origin for the Julien *et al.* model [1] (dashed curve), and a local moment model with a small induced spin density oscillation (solid curve) as described in the text. The data is a one dimensional representation of the absolute value of $<s>$. (b) Plot of simulated $^{89}$Y NMR data from YBa$_2$(Cu$_{1-x}$Zn$_x$)$_4$O$_8$ with x=0.005 at 50 K and 14 T expected for the model of Julien *et al.* (dashed curve [1]) and the local moment model with a small induced spin density oscillation (solid curve). The spatial dependence of $<s>$ in both cases is plotted in (a). (c) Plot of the corresponding simulated $^{63}$Cu NMR data using the Julien *et al.* model (dashed curve [1]) and the local moment model with a small induced spin density oscillation (solid curve).